\documentclass[11pt]{article}
\usepackage{moriond,epsfig}

\usepackage{color}
\usepackage{braket}
\usepackage{amsmath,amssymb,amsfonts}
\def\be{\begin{equation}}
\def\ee{\end{equation}}
\def\bea{\begin{eqnarray}}
\def\eea{\end{eqnarray}}

\newcommand{\eq}[1]{Eq.~\ref{#1}}
\newcommand{\Figref}[1]{Fig.~\ref{#1}}

\newcommand{\nn}{\nonumber}

\def\bea{\begin{eqnarray} }
\def\eea{ \end{eqnarray} }

\begin{document}
\vspace*{4cm}
\title{GLUINOS LIGHTER THAN SQUARKS AND DETECTION AT THE LHC}

\author{ L. Velasco-Sevilla}

\address{CINVESTAV-IPN, Apdo.~Postal 14-740, 07000, M\'exico D.F., M\'exico}

\maketitle

\abstracts{We explain how physics in the kaon sector proves useful in setting constraints in models where all the supersymmetric particles, except the gluino, are in the range  $\sim(3-10)$ TeV. We also discuss the signals of these models at colliders, in particular at the LHC. }

\section*{Introduction }
Here we discuss two important features of scenarios where there are three families of heavy squarks, while the mass of the gluino is considerable lighter than the rest of the other supersymmetric particles. The two features we  discuss are: (i) bounds that can be obtained from flavour changing neutral currents (FCNC) in the kaon sector and (ii) the signatures of these models at colliders. In section I we disambiguate our scenario from others existing in the literature. In section II we present the restrictions from kaon physics, that can prove useful in setting some bounds of these models, together with typical conditions set by an underlying family symmetry. Then in section IV, we discuss the signatures at colliders, specifically at the LHC.

\section{Models with lighter gluinos than squarks\label{sec:models}}

Since there are many models in the literature with  gluinos lighter than squarks, we disambiguate the scenario that we consider here with those in the literature and specify here our interest on it. Gauge Anomaly Mediation Symmetry Breaking (GAMSB) scenarios set soft terms, at tree level, equal to zero at a high-scale before the symmetry is broken. Then, loop corrections induce non zero masses and, in particular, the masses of the gauginos, $M_i$, $i=1,2,3$, are different (e.g. \cite{Gherghetta:1999sw}). On the other hand, G2-MSSM models, based on supergravity, achieve at $M_P$, $m_0=m_{3/2}=A_0$, while also achieving a split among the gaugino masses (e.g. \cite{Acharya:2008hi}). Both models, however, typically have the common feature that
$M_2> M_1 > \ M_3$, 
which sets a lower mass to the gluino, while leaving heavier scalars. How large is the split between the squarks and the gluino, depends on the details of the particular models. However, in fact this kind of spectra is not difficult to obtain from effective supergravity scenarios coming from string  compactifications where the overall modulus, and not  the dilaton, gives the main contribution to gaugino  masses \cite{Brignole:1997dp}. On the other hand, generic {\it split supersymmetry} models \cite{Cohen:1996vb} set, as a working condition, a split between scalars and fermions, but one has  some freedom in setting the precise split, as long as an agreement with all the observable physical quantities is satisfied.  Models based on an underlying $U(2)$ family symmetry (e.g. \cite{Barbieri:2011ci}) achieve a split between the masses of two families of heavier squarks and one lighter, together with a split of the gluino mass, which is also light. 

The scenario that we consider here sets the following ranges for the masses of the gluino and the squarks, respectively,
\bea
\label{eff:conditionsmass}
 900 \   {\rm{GeV}} \  \leq m_{\tilde g}  \leq  2000 \ {\rm{GeV}},\quad
3000 \   {\rm{GeV}} \  \leq m_{\tilde q}  \leq  10000 \ {\rm{GeV}}.
\eea
Spectra of this type agrees with all the observable physical quantities  if $4 \lesssim \tan\beta \lesssim 20$.
 The exact details of achieving electroweak symmetry breaking conditions require  a small fine tuning, as is expected as soon as one starts increasing the scale of some supersymmetric particles. However, as mentioned before, these models could arise from some supergravity scenarios, so if the conditions in \eq{eff:conditionsmass} are a result from boundary conditions at a high-scale, the fine-tuning would be controlled by the underlying scenario.

\section{Limits from kaon physics}

It is well  known that if physics,  beyond the SM, in the kaon sector would contribute at tree level, then the effective hamiltonian measuring the $\Delta S=2$ transitions would require a new physics scale of about $\Lambda=10^{4}$ TeV. In the general MSSM, the contribution to  $\Delta S=2$ comes at the loop level, so the scale decreases, nevertheless  it could  be used to put constraints of FCNC processes from scenarios where only some particles   are heavier ($O(10)$ TeVs) and other are much lighter. The values of $\Delta m_K$ and the flavour violating parameter $\epsilon_K$ can be used to set limits on the overall scale of the mass of the scalars and of course in the off-diagonal masses.  $\Delta m_K$  is proportional to the effective Hamiltonian of the $\Delta S=2$ transitions, $\Delta m_K =2 {\rm{Re}}\braket{  K^0|H_{\rm{eff}}^{\Delta S=2} |\overline{K}^0}$, while  $\epsilon_K$ is proportional to the imaginary part, $|\epsilon_K|=$ $|{\rm{Im}} \braket{  K^0|H_{\rm{eff}}^{\Delta S=2} |\overline{K}^0}|/(\sqrt{2}\Delta m_K)$.  
The values of  $\Delta m_K$ experimentally and in the SM  are respectively
\bea
\label{eq:}
\Delta m_K =(3.483\pm 0.0059) \times 10^{-15},\quad
\Delta m^{\rm{NNLO}}_{K \rm{(SD)}}=(3.1 \pm 1.2)\times 10 ^{-15} \ {\rm{GeV}},
\eea
which we can see that they agree at the one sigma C.L. so the use in physics beyond the SM is to set limits in the contributions that the new processes can give. On the other hand, it is well known that long distance effects contributing to $\Delta m_K$  cannot be precisely calculated, so the best we can do is to use the most precise values of the short distance effects, which we have denoted as $\Delta m^{\rm{NNLO}}_{K\rm{(SD)}}$ and whose  value has been calculated in \cite{Brod:2011ty}. Despite  the uncertainties due to the unknown long distance effects, one can still use $\Delta m_K$  to set limits on the masses of the squarks because for generic values of  $(\delta^d_{XY})_{ij}$, $X,Y=L,R$ (left, right), the contributions to $\Delta m_K$ can easily overcome values of $\times 10 ^{-15} $. So effectively what we can do is to use use then $\Delta m^{\rm{NNLO}}_{K \rm{(SD)}}$ and its corresponding uncertainty as
$|\Delta m_K^{\rm{S}}| < |\overline{\Delta m_K^{\rm{exp}}} - \overline{ \Delta m_{K(SD)}^{\rm{NNLO}}}   +2 (\sigma_{\Delta m_{K(SD)}^{\rm{NNLO}}} +   \sigma_{\Delta m_K^{\rm{exp}}}) |$, to set a limit on the supersymmetric contribution, $ \Delta m_K^{\rm{S}}$. The values of  $\epsilon_K$ experimentally and in the SM \cite{Brod:2011ty} are respectively
\bea
|\epsilon_K|= (2.228 \pm 0.11)\times 10^{-3},\quad
|\epsilon^{NNLO}_K|= (1.81 \pm 0.28 ) \times 10^{-3} ,
\eea
although these values also agree within the $2\sigma$ C.L., and there  are many uncertainties in all the quantities involved in the calculation of  $|\epsilon_K|$ (\cite{Brod:2011ty} and Hoelbling's talk), we can use them to constraint the contributions from supersymmetric processes, because just as in the case of $\Delta m_K$, they can easily overcome the experimental value.  This time, however, we could still hope that supersymmetric contributions would prove useful in bringing closer the SM and the experimental values. The effective Hamiltonian, $H^{\rm{eff}}=\sum_i^{5} C_i O_i\eta_i +\sum_i^{3} \tilde C_i \tilde O_i\eta_i $ {\footnote{For the exact expressions of all the quantities mentioned here, please check \cite{Kadota:2011cr}. The details of the QCD evolution is going to be presented in \cite{Kersten2012wil}.}}, depends, through the Wilson coefficients $C_i$  only on the  flavour violating parameters $(\delta^d_{XY})_{12}$, which can be defined as $(\delta^d_{XY})_{ij}$ $=(\hat m^{d 2}_{XY})_{ij}/\sqrt{\hat m^{d 2}_{XX})_{ii}  \hat m^{d 2}_{YY})_{ij} }$, where $\hat m^{d 2}_{XY}$ represent the effective mass terms appearing in the effective $6\times 6$ squared mass matrix giving masses to the six d squarks, in the basis where Yukawa couplings are diagonal. For the parameters $(\delta^d_{LR})_{12}$, we have
\bea
\label{eq:delta12d}
(\delta^d_{LR})_{12}=\frac{-a^d_{12}v_d}{\sqrt{\hat m^{d 2}_{LL})_{11}  \hat m^{d 2}_{RR})_{22} }}
\approx
\frac{A_d m_b}{m_{3/2}^2} \epsilon^r\leq 10^{-2},
\eea
where of course the term proportional to the $\mu$ term disappears since it is proportional to the Yukawa coupling. The term proportional to the trilinear term $a^d=Y^d A^d$ can still be off-diagonal because proportionality to Yukawa couplings is just a boundary condition at $M_P$ and whenever the  terms $A_f\neq 0 $, then $a^d$ runs different than $Y^f_{ij}$.  The parameter $\epsilon$ in 
 \eq{eq:delta12d}  refers to a FS parameter and so depending on the value of $r$, they can saturate  the value of $O(10^{-2})$. To get this maximum value, $O(1)$ parameters in going from the third expression after the initial $=$ sign were taken into account.
 The parameters  $(\delta^d_{LL})_{12}$ and $(\delta^d_{RR})_{12}$ cannot be expressed as $(\delta^d_{LR})_{12}$ because they are  more model dependent than it, however in FS both, imaginary and real parts,  typically do not exceed $O(10^{-1})$. 
 
\begin{figure}
\centering
\includegraphics{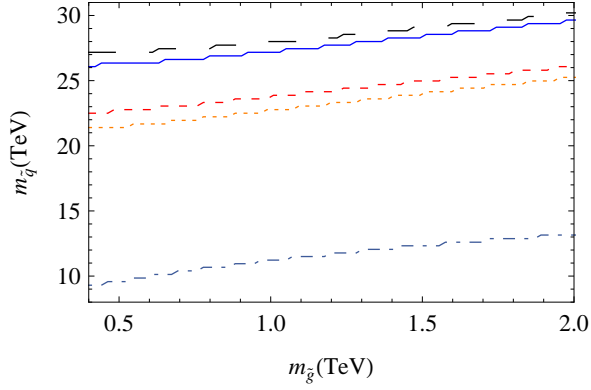}
\caption{Limits on the mass of the squarks of the type d, $m_{\tilde{q}}$,  as a function on the mass of the gluino, $m_{\tilde g}$, for a case where only
 $\sqrt{| \rm{Re}( (\delta^d_{LL})_{12}   (\delta^d_{RR})_{12} ) |}$ contributes, with a value equal to $0.1$. The solid (blue) continue line corresponds to the NLO evolution, while the medium dashed (red) line corresponds to the LO approximation. Note that the difference between NLO and LO is of the $O(15\%) $, which proves the importance of the QCD NLO determination for these cases of heavy squarks. 
For the rest of the lines, please refer to the text.
\label{fig:QCDeffmass}}
\end{figure}

Another important motivation for this work is to reassess the importance of QCD contributions. Naively, one would think that at few TeVs, LO corrections would be enough, however as it was pointed out in \cite{Contino:1998nw}, in the context of two heavy families of squarks, NLO effects can still have an important impact. In this context, where the three families  of squarks are of the same order of magnitude, the case is not different.   Then, according to the motivations above, what  do here, as a complement of  \cite{Kersten2012wil}, is  to 
 \begin{enumerate}
 \item Assess the importance of considering NLO effects in setting the limits of the masses of d type squarks and the flavour violating parameters $(\delta^d_{XY})_{12}$ for the case of three heavy squarks families and a light gluino.
 \item Check that  the values of squarks masses, in the range considered, and the real parts of $(\delta^d_{XY})_{12}$ satisfy the bounds set by $\Delta m_K$.
 \item Put constraints on the imaginary parts of  $(\delta^d_{XY})_{12}$.
 \end{enumerate}
  In \Figref{fig:QCDeffmass} we show a plot of the mass of the gluino $m_{\tilde{g}}$ versus the common mass scale of the three heavy  squarks families. In this plot we have included:  (i) a solid (blue) continue line corresponding to  the NLO evolution, (b) a medium dashed (red) line corresponding to the LO approximation, (c) a small-dashed (orange) light line corresponding to the  LO  evolution using VIA approximation for bag parameters and (d) a dot-dashed (light blue) line corresponding to the results without QCD corrections. For comparison, we have additionally plotted a medium-dashed dark (black) line representing the NLO evolution of the case where only two families where decoupled at the scale $m_{\tilde q}$, one is light and we have followed a similar QCD evolution to the three heavy family case.  Apart from showing the importance of NLO QCD corrections, the curves in \Figref{fig:QCDeffmass} show that for the scenarios in the present discussion, the proper QCD evolution, sets a safe limit on the masses of the squarks of the type d considered.

\begin{figure}
\includegraphics{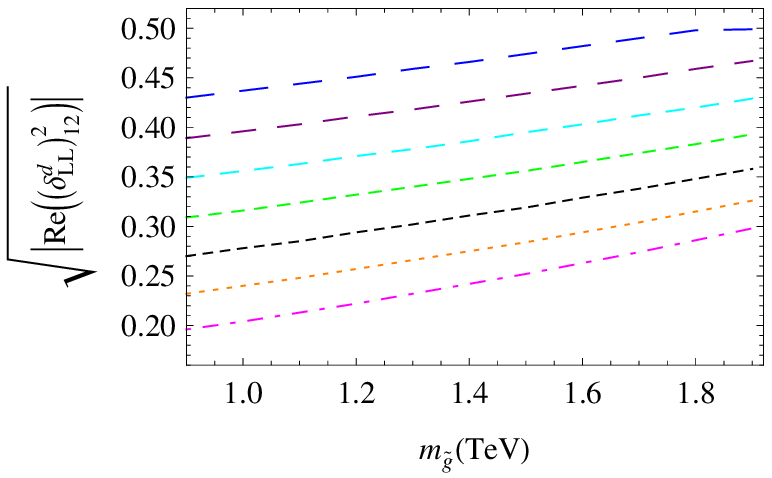}
\includegraphics{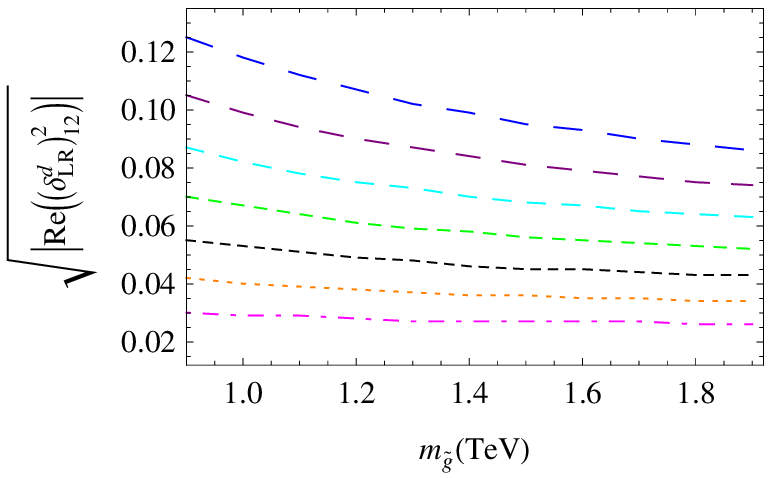}
\includegraphics{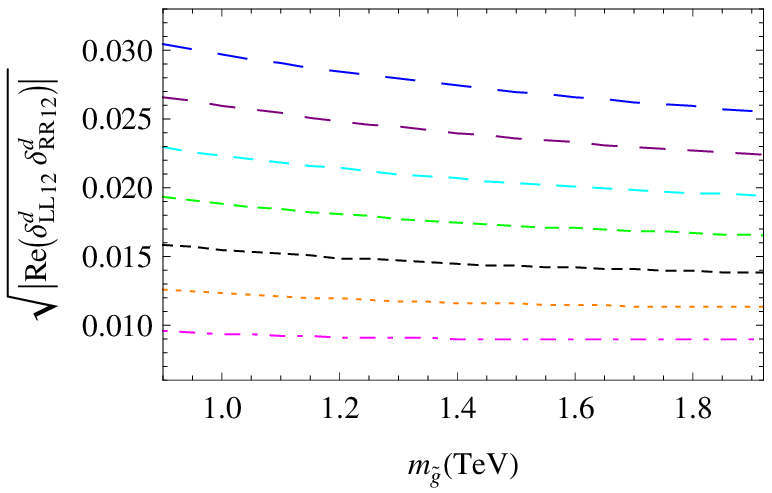}
\includegraphics{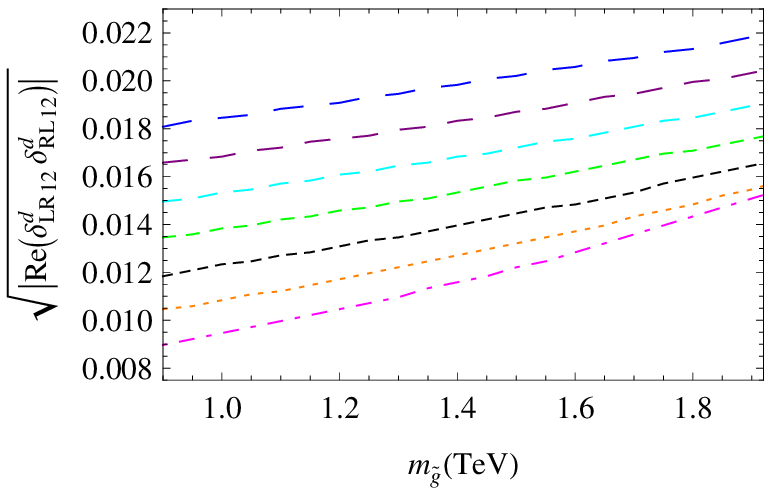}
\caption{Limits on $\sqrt{|{\rm{Re}}( ((\delta^d_{XY})_{12})   ((\delta^d_{X'Y'})_{12})  ) }|$, using $\Delta m_K$.  In all cases only the flavour violating parameters shown in the vertical axis have been set different to zero. The limits on $\sqrt{|{\rm{Re}}[(\delta^d_{RR})^2_{12}]|}$ and  $\sqrt{|{\rm{Re}}[(\delta^d_{RL})^2]|}$ can be obtained by interchanging $R\leftrightarrow L$ in the upper two panels, respectively. The two lower panels show, respectively, the results for  $\sqrt{|{\rm{Re}}[\delta^d_{LL} \delta^d_{RR}]_{12}}|$ and $\sqrt{|{\rm{Re}}[\delta^d_{LR} \delta^d_{RL}]|_{12}}$.
 The  different curves correspond, from bottom to top to  $m_{\tilde q}=4,\hdots , 10$ TeV.
\label{fig:Redeltas}
}
\end{figure}

\begin{figure}
\includegraphics{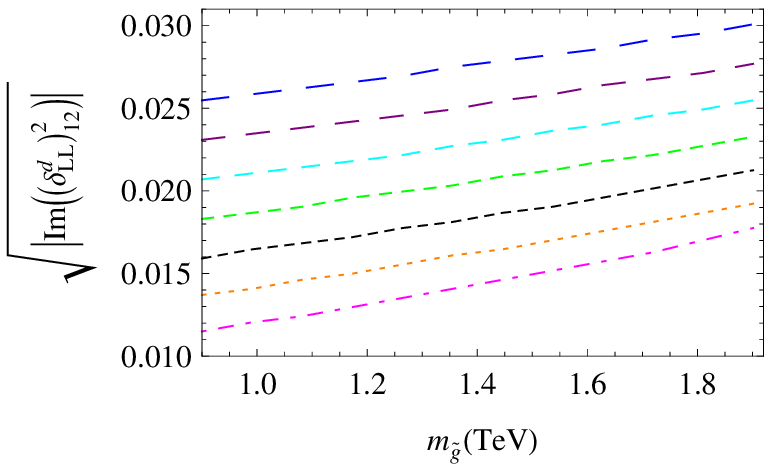}
\includegraphics{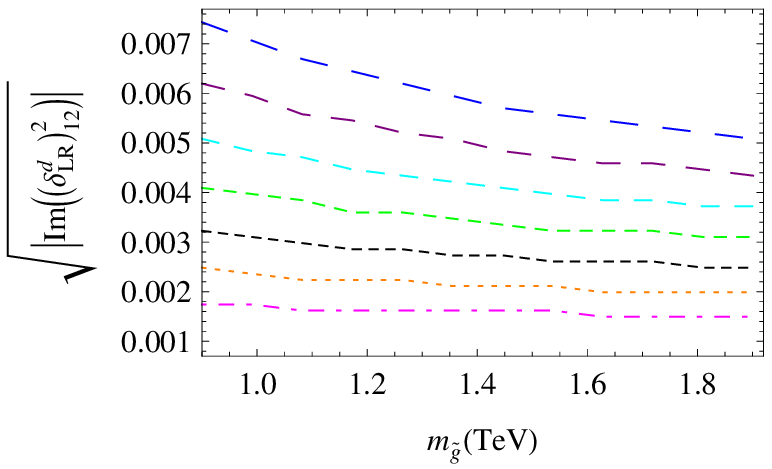}
\caption{Limits on $\sqrt{|{\rm{Im}}[(\delta^d_{XY})_{12} (\delta^d_{X'Y'})_{12}   ]|}$, using $\epsilon_K$.   The limits on $\sqrt{|{\rm{Im}}[(\delta^d_{RR})^2_{12}]|}$ and  
$\sqrt{|{\rm{Im}}[(\delta^d_{LR})^2_{12}]|}$ can be obtained by interchanging $R\leftrightarrow L$,  respectively. Limits for the  case where  $\sqrt{|{\rm{Im}}[(\delta^d_{LL})_{12} (\delta^d_{RR})_{12}]|}$ and $\sqrt{|{\rm{Im}}[(\delta^d_{LR})_{12} (\delta^d_{RL})_{12}]|}$,  just as in the real case, are  approximately one order of magnitude smaller than the corresponding case where only one of either  $\delta^d_{RR}$ or  $\delta^d_{LL}$ are present. 
The  different curves correspond, from bottom to top, to  $m_{\tilde q}=4,\hdots, 10$ TeV.
\label{fig:Imdeltas}}
\end{figure}

In \Figref{fig:Redeltas} we present curves for  fixed values of $m_{\tilde q}$, plotting $m_{\tilde g}$ against the limits on 
 $\sqrt{| \rm{Re}( (\delta^d_{XY})_{12}   (\delta^d_{X'Y'})_{12} ) |}$.
 Note then that, according to \eq{eq:delta12d}, at the scale of squarks of the type d of  $\leq O(10)$ TeV, the use of $\Delta m_K$ proves relevant since for some cases, the limit set by it,  is in fact close to the allowed value for $(\delta^d_{RR})_{12}$.
 The same happens when contributions from $(\delta^d_{RR})_{12}$ are present. For other cases, where only $(\delta^d_{LR})_{12}$ and/or $(\delta^d_{RL})_{12}$ contribute, we can check that these models give non-dangerous contributions to $\Delta m_K$.
Finally, in \Figref{fig:Imdeltas}  we present curves for fixed values of $m_{\tilde q}$, plotting $m_{\tilde g}$ against the limits on $\sqrt{|{\rm{Im}}( ((\delta^d_{XY})_{12})   ((\delta^d_{X'Y'})_{12})  ) }|$.
As expected, the limits on these quantities, set by $\epsilon_K$, prove to be indeed relevant, since the limits lie above the maximum values that in principle FS can achieve, so these limits are a useful constraint for these scenarios.

\section{Signatures at the LHC}

This scenario features the same kind of signatures at colliders, as the models cited in section \ref{sec:models}, which are mainly the following:
\begin{enumerate}
\item $\Delta m_\chi=m_{\tilde\chi_1^{\pm}}-m_{\tilde\chi_1^{0}}=O(100)$ MeV, where $m_{\tilde\chi_1^{\pm}}$ and $m_{\tilde\chi_1^{0}}$ are, respectively, the masses of the lightest chargino and the lightest neutralino.  The order of magnitude of  $\Delta m_\chi$ leads to the existence of soft pions.
\item  4 SM fermion final states as a result of the gluino decays
\small
\bea
&&pp \rightarrow \quad \tilde g\tilde g,\quad \tilde \chi^0 _1\tilde \chi^{\pm} _1,\quad \tilde \chi^{\pm}_1\tilde\chi^{\mp}_1\nn\\
&&\begin{array}{ll}
 \tilde g \quad  \rightarrow   \tilde \chi^0_2 \ t \bar{t} &  {\rm{Decays\  of \ the \ secondary\  charginos/neutralinos}}  \\
            \quad   \rightarrow   \tilde \chi^0_1\ b \bar{b}  &  \tilde \chi^0_2  \ \rightarrow  \  \tilde \chi^\pm_1 W^* \  \rightarrow  \   \tilde \chi^\pm_1 \ell \nu_{\ell},\quad \tilde \chi^\pm_1 q q^\prime  \\
            \quad   \rightarrow   \tilde \chi^0_1 \ q \bar{q} &  \tilde \chi^\pm_1  \ \rightarrow  \    \tilde \chi^0_1    \  \rightarrow  \    \tilde \chi^0_1   \ell \nu_{\ell},\quad
     \tilde \chi^0_1 q q^\prime\\
            \quad   \rightarrow   \left(\tilde \chi_1^- \bar{d}u + h.c.\right). &
\end{array}            
\eea
\item Effective supersymmetric cross section at $\lesssim 2$ TeV, produced mainly by gluinos, that could be easily tested at the LHC.
\end{enumerate}
The first of the items above occurs  whenever there is a large $|\mu|$ parameter. At zeroth order, in fact $m_{\tilde\chi_1^{\pm }}^{(0)}-m_{\tilde\chi_1^{0}}^{(0)}$=$M_2$, at the first loop level  they receive corrections from mixing with the Higgsino,  then, at  next order, there will be a mixing with the $U(1)_Y$ gaugino, enhanced by $\tan\beta$ and dependent on the difference of $M_1$ and $M_2$:
$
m_{\tilde\chi_1^{\pm }}^{(2)}-m_{\tilde\chi_1^{0}}^{(2)}$  $=
\frac{m^4_W\tan^2 \theta_W}{(M_1-M_2)\mu^2} \sin^2 2\beta.
$
This condition produces the decay of the charginos into neutralinos and soft-pions, $\chi^{\pm}\rightarrow \chi^0 \pi^{\pm}$, then there would be displaced vertices with a track of  few centimeters long. However, current strategies of long-lived charged particles that make it all the way through the detector seem to evade the detection of the kind of decays generated by this kind of neutralinos (e.g. \cite{Torro:2009zz}). However, if experimental techniques can overcome the difficulties in detection, this could provide indeed an interesting discerning criteria for these models.  For systematic studies of this signature, although  in different models than the present, see for example
\cite{Chen:1996ap}. The second of the signals above, although not unique for the scenario discussed here, are interesting because they clearly signal a split of the supersymmetric scalars. The difficulty in detection of course arises from the fact that it would be difficult at the LHC to disentangle the QCD background from a possible signal. Finally, for the third  of the items above, LHC is already setting limits on what the  effective supersymmetric cross section of related models should be. We find that a typical supersymmetric cross section, produced only by gluinos of $m_{\tilde g}=650$ GeV, is $\sigma_{\rm{SUSY}}\approx 0.2$ pb,  at a luminosity $L\leq 800 {\rm{pb}}^{-1}$, and an energy, at the center of mass, of $\sqrt{7}$ TeV. For  gluinos of $m_{\tilde g}=850$ GeV, $\sigma_{\rm{SUSY}}$ decreases to $\approx 0.04$ pb, to finally decrease  to $\sigma_{\rm{SUSY}}\approx 0.001$ pb for  $m_{\tilde g}=1$ TeV (check \cite{Feldman:2010uv} for similar results in the context of the G2-MSSM models).  In fact, ATLAS and CDF have already put limits on related scenarios, where they set the mass of all the supersymmetric particles at $10$ TeV, except of course the gluino mass. They have improved the previous exclusion of 2010 of gluinos with  $m_{\tilde g}=580$ GeV. In particular, ATLAS (see M. D. Joergensen talk and \cite{atlas:2012}) has determined that  gluinos with a mass lower than $m_{\tilde g}=810$ GeV are excluded. As explained before, the scenario considered here, sets the values of the squarks of the type d below, so the limits on the gluino masses from the experimental cross sections already determined, serve just as an illustration of the capacity of the LHC in setting bounds for the gluino mass. A detailed study at the LHC will require specific details of the kind of models presented here.

\section*{Acknowledgments}
I thank J. Kersten for collaboration leading to some of the results presented here and the organizers of the 47th Recontres de Moriond for such a pleasant conference.

\section*{References}


\end{document}